\documentclass[aps,pre]{revtex4}
\usepackage{epsfig}
\begin{document}

\title{Multifractal Collision Spectrum of Ballistic
Particles with Fractal Surfaces }

\author{S. B. Santra$^{1,2}$ and  B. Sapoval$^{1}$}.

\affiliation{$^{1}$ Laboratoire de Physique de la Mati\`ere
Condens\'ee, Ecole Polytechnique, 91128 Palaiseau C\'edex, France. \\
$^{2}$Department of Physics, Indian Institute of Technology Guwahati,
Guwahati-781039, Assam, India.}

\date \today

\begin{abstract}
Ballistic particles interacting with irregular surfaces are
representative of many physical problems in the Knudsen diffusion
regime. In this paper, the collisions of ballistic particles
interacting with an irregular surface modeled by a quadratic Koch
curve, are studied numerically. The $q$ moments of the source spatial
distribution of collision numbers $\mu(x)$ are characterized by a
sequence of ``collision exponent'' $\tau(q)$. The measure $\mu(x)$ is
found to be multifractal even when a random micro-roughness (or random
re-emission) of the surface exists. The dimensions $f(\alpha)$,
obtained by a Legendre transformation from $\tau(q)$, consist of two
parabolas corresponding to a trinomial multifractal. This is
demonstrated for a particular case by obtaining an exact $f(\alpha)$
for a multiplicative trinomial mass distribution. The trinomial nature
of the multifractality is related to the type of surface
macro-irregularity considered here and is independent of the
micro-roughness of the surface which however influence the values of
$\alpha_{min}$ and $\alpha_{max}$. The information dimension $D_I$
increases significantly with the micro-roughness of the
surface. Interestingly, in contrast with this point of view, the
surface seems to work uniformly. This correspond to an absence of
screening effects in Knudsen diffusion.

\end{abstract}

\maketitle

\section{Introduction}
Multifractals appear in a wide range of situations like energy
dissipation in turbulent flows \cite{turbl}, electronic eigenstates at
metal insulator transition \cite{mit}, fluctuations in finance
\cite{finc}, dynamics of human heartbeat \cite{hhb} and many
others. Recently, multifractal behavior of the first passage time has
been studied in the transport of diffusing particles on a $2D$
Sierpinski gasket with absorbing and reflecting barriers
\cite{kim}. Our interest is to investigate the multifractal properties
associated with the collision spectrum of ballistic trajectories of
point particles interacting with irregular surfaces. 

The emergence of the concept of fractal geometry has provided an
efficient tool to model the influence of strong geometrical
irregularity on various physical and chemical processes
\cite{mandelsapo}. Ballistic trajectories in systems with irregular
geometry are representative of several physical situations. For
example, they correspond to the path of light rays in irregular
structures when diffraction phenomena can be neglected. They also
represent the path of atoms or molecules in confined vessels at
sufficiently low pressure, $i.e.$; in the Knudsen diffusion regime
\cite{knud}. The interaction of atoms or molecules with irregular
surfaces plays a major role in heterogeneous catalysis \cite{catlys}
where porous catalysts have often very irregular surfaces down to the
molecular scale. Due to the geometrical irregularity, the small scale
structure may confine reactants and increase the interaction with the
surface. Xenon nuclear magnetic resonance is used to probe porous
structures through the interaction of rare gas atoms with the solid
surfaces \cite{levitz}. The properties of ballistic trajectories in
$2D$ pre-fractal surfaces has already been studied in connection with
the chaotic aspects, possibility of ``open'' black body, Knudsen
diffusivity, mean interaction, and catalytic efficiency by Santra et
al\cite{santra}. Coppens and co-workers have also investigated the
influence of the $3D$ fractal surface morphology on Knudsen diffusion
in nanoporous media \cite{coppens}. Multifractality has already been
observed in the molecular diffusion on random fractal structures with
time-dependent random potential \cite{mdiff}.

The interest here is to explore the multifractal aspects of the number
of collisions made by $2D$ ballistic trajectories within irregular
surfaces in the Knudsen diffusion regime. The spatial distribution of
the collision numbers of these ballistic trajectories over the slit
source is found to be multifractal.

\section{The Model}
\begin{figure}
\bigskip
\centerline{\hfill \psfig{file=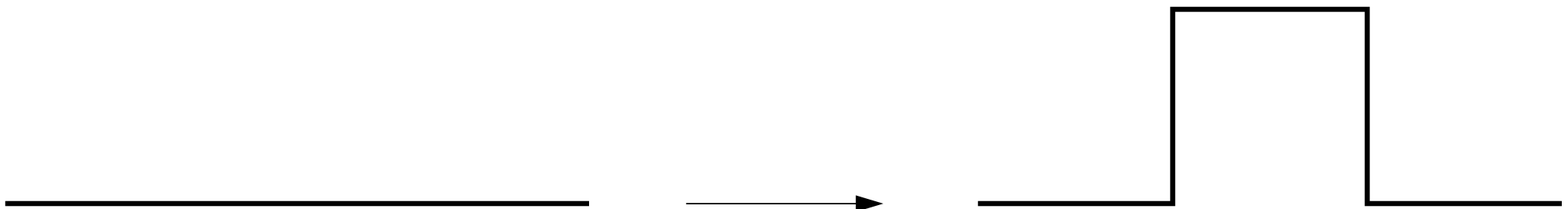,width=0.3\textwidth} \hfill}
\centerline{(a) Pre-fractal generator}
\bigskip
\centerline{\hfill \psfig{file=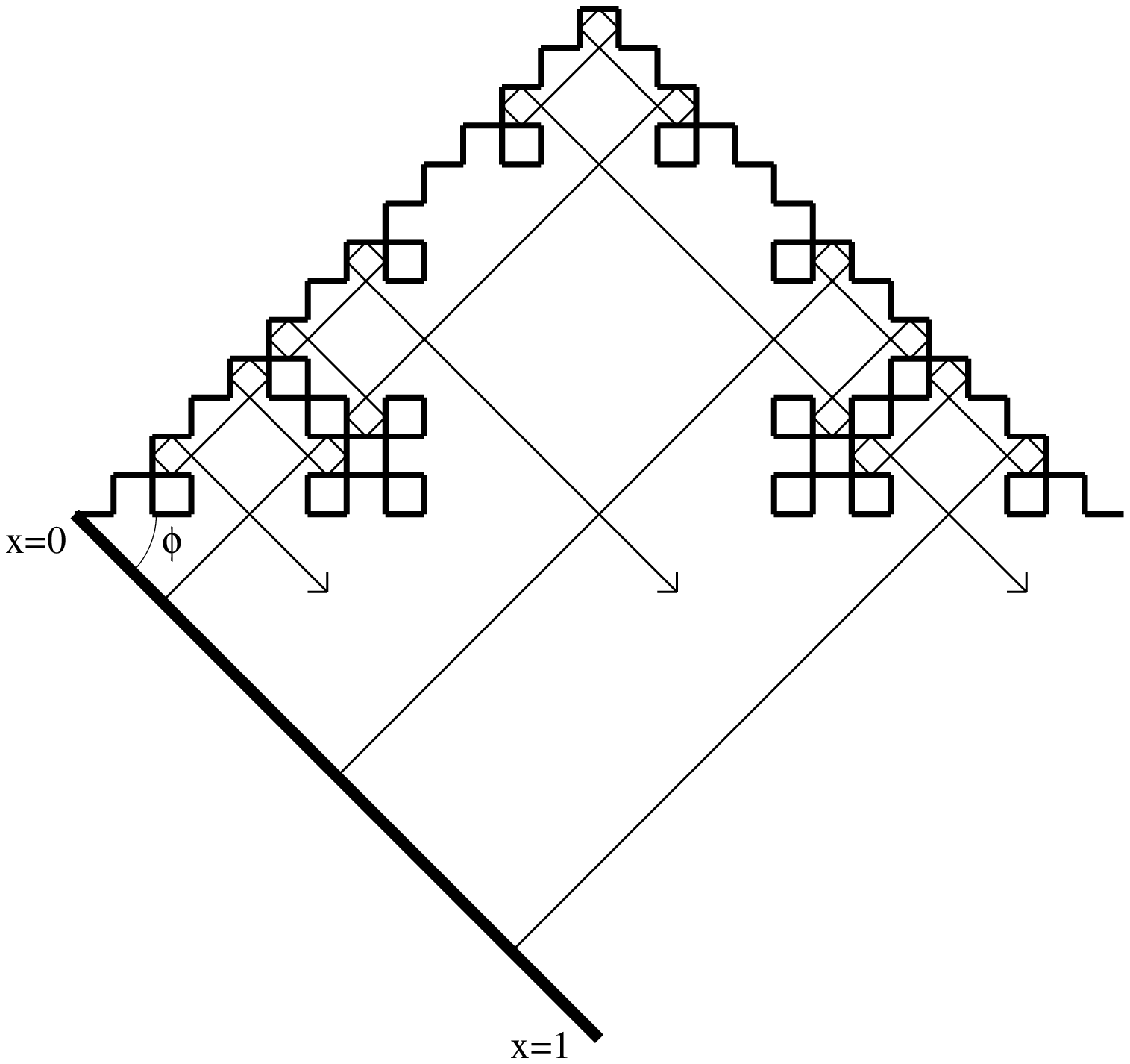,width=0.3\textwidth} \hfill
\psfig{file=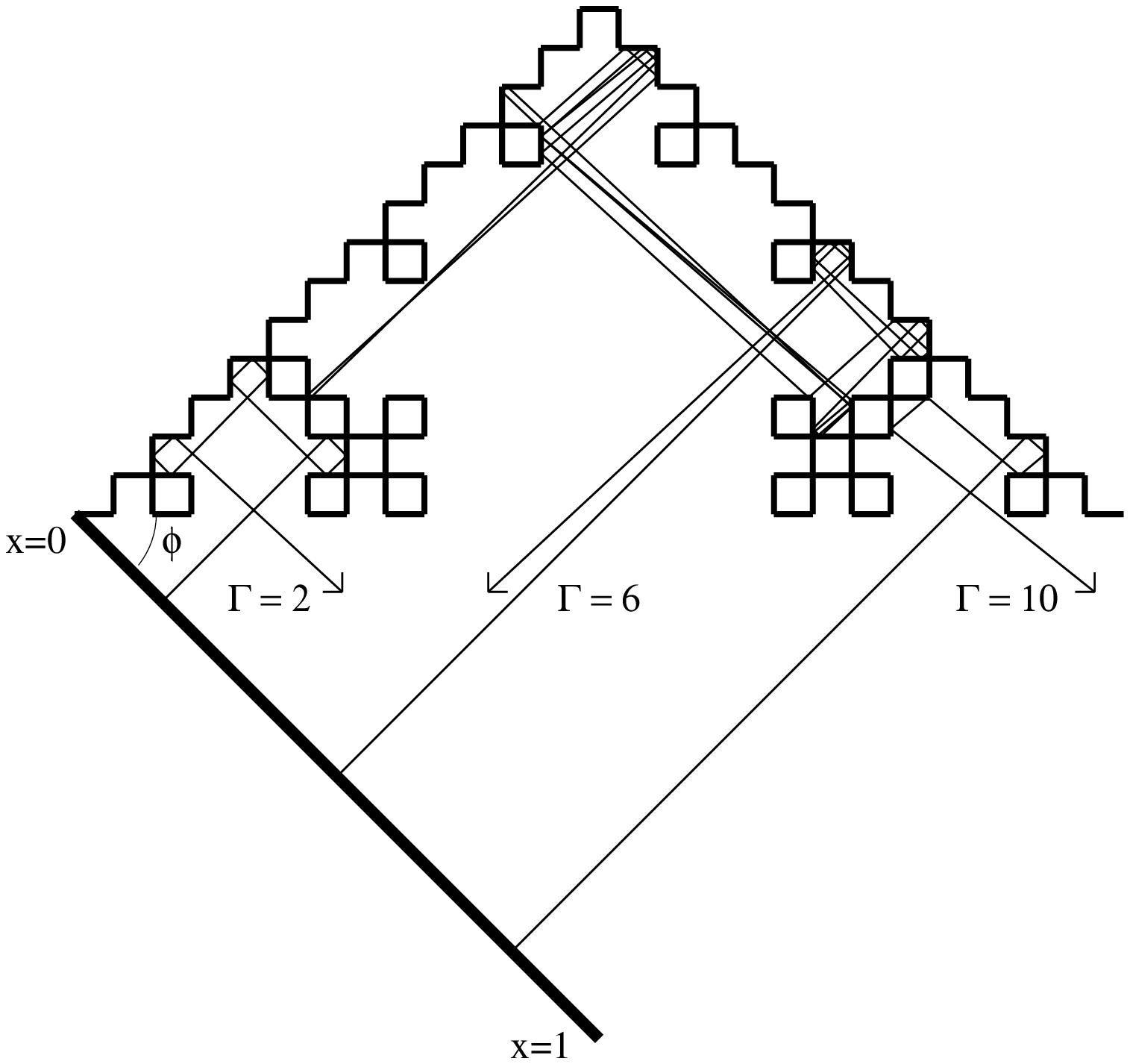,width=0.3\textwidth}\hfill}
\centerline{\hfill (b) \hfill\hfill (c)\hfill}
\bigskip
\caption{\label{demo1} Schematic representation of the collision
process at the third generation of the pre-fractal surface, a
quadratic Koch curve of dimension $\ln5/\ln3$. The generator of the
pre-fractal is shown in ($a$). The thick lines, extended from $x=0$ to
$x=1$, in ($b$) and $(c)$ represent the slit. The slit angle $\phi$ is
$45^\circ$ here. In ($b$), the trajectories of particles with specular
reflections form three different position on the slit are shown.  The
central trajectory makes $3^3=27$ collisions and the other two make
$9$ collisions each. The trajectories of non specular reflection are
shown in ($c$). These trajectories are generated for $\Gamma=2$, $6$
and $10$ for the same initial positions $x$ and slit angle $\phi$ of
($b$). }
\end{figure}

The macro-irregularity of the surface is modeled by a quadratic Koch
curve of fractal dimension $\ln5/\ln3$. The pre-fractal generator is
shown in Fig.\ref{demo1} $(a)$. Particles are launched from a slit of
unit length, placed at an angle $\phi$ with the zeroth order of the
pre-fractal.  The particle trajectories due to specular collisions
within the third ($\nu=3$) generation of the pre-fractal are shown in
Fig.\ref{demo1} $(b)$. Solid lines with an arrow at the end represent
the trajectories. The slit source is represented by a thick line
extended from $x=0$ to $x=1$ placed at $\phi=45^\circ$ with the zeroth
stage of the pre-fractal surface, a horizontal line.

A multifractal measure is in general related to the distribution of a
physical quantity on a geometrical support \cite{feder}. Here, the
geometrical support is the slit source from which the trajectories are
launched and the multifractal measure $\mu(x)$ is the distribution of
the relative collision numbers, ratio of the number of collisions made
by a particle to the total number of collisions made by all the
particles, over the slit. The particles leave the slit normally. Each
trajectory is followed as long as they are within the pre-fractal
surface. A particle trajectory is described by the successive
collisions with the surface. The collision points are determined by
solving the linear equations describing the particle trajectory and
the elements of the pre-fractal surface. The trajectories in
Fig.\ref{demo1}$(b)$ are generated using this algorithm.

For the pre-fractal $\nu$th generation, the slit is divided into
$3^\nu$ equal segments of length $1/3^{\nu}$ each. A particle is
placed at the center of each small part. There are then $3^\nu$
particles launched from the slit. The number of collisions $n(x)$ made
by a particle initiated from the position $x$ is counted and the
relative collision number $\mu(x)$, ratio of $n(x)$ to the total
number of collisions $\sum_x n(x)$ made by all the particles, is then
calculated. The value $\mu(x)$ is assigned to the slit segment
positioned at $x$.

Generally, particle reflections with surfaces at finite temperatures
are non specular and the angle of reflection, or re-emission
$\theta_r$ is different from the angle of incidence $\theta_i$.  Due
to the interaction with the surface, the angle of reflection has a
distribution around the angle of incidence. Also the reflection angle
could be arbitrary due to the micro-roughness of the
surface. Non-specular reflection for atoms or molecules may come from
the complex nature of the particle-wall interaction \cite{valleau}. At
the time of hitting the surface, the particle may stay adsorbed for
some time and then may desorb, exchanging momentum with the
surface. For this reason, the particle trajectories as a whole have a
random character. In order to mimic the random character of each
collision, the angle of reflection $\theta_r$ is supposed to be
related to the angle of incidence $\theta_i$ as
\begin{equation}
\label{angle}
\theta_r=\theta_i+\delta\theta
\end{equation}
where $\delta\theta$ is a random angle distributed uniformly over a
range of angle $\pm\Gamma^\circ$. This range describes qualitatively
the random character of the collisions and $\Gamma$ can also be called
the surface micro-roughness.  In the following, the $\Gamma$ values are
expressed in units of degree. A few non-specular trajectories computed
using the above rule, are shown in Fig.\ref{demo1}$(c)$.

In the case of specular reflections, the reflection angle is not
determined from numerical computation of trigonometric functions since
the elements of the pre-fractal geometry are just parallel to the $x$
and $y$ axes. Consequently, there is no numerical error due to
truncation of angle computations. The only truncation errors come from
the solution of the linear equations. Because of this error, there
could exist trajectories, besides real ray splitting between close
trajectories hitting salient corners, which are modified by a
``spurious'' splitting due to number truncation. To avoid spurious
beam splitting, the cumulative error on the point of collision due to
the numerical precision of the computer is calculated. For specular
reflections, the initial positions of the particles are chosen in such
a way that this error is always less than the distance between the
point of collision and the nearest corner. For non-specular
reflections, those trajectories are rejected and the particles are
resent from the same position with a new random number sequence.

It should be emphasized that for $\phi=45^\circ$, the number of
specular reflections for the particular Koch curve considered here can
be determined from a simple recursion relation without solving the
linear equations for the trajectory and the surface elements. The
number of collisions up to the third generations of the pre-fractal
surface are listed in Table $1$. In the first generation, all the
particles send from the central part of the slit have $3$ collisions
whereas the particles from the other two parts have only $1$ collision
each. The collision numbers are then in the form of $(1$,$3$,$1)$ in
the first generation. In the next pre-fractal generation, each part of
the the slit is again subdivided into another $3$ parts. It can be
seen that for $\nu=2$, the collision numbers $(3$,$9$,$3)$ for the
particles launched from the central part of the slit is just
$3\times(1$,$3$,$1)$ where $3$ is the number of collisions
corresponding to the same part of the slit for the previous
generation. The other two sequences $(1$,$3$,$1)$ for the two other
parts of the slit are given by $1 \times (1$,$3$,$1)$. The
trajectories from these two parts had a single collision each at the
previous generation. The multiplication of the collision numbers from
the first to the second generation of the pre-fractal is shown in
Fig.\ref{demoe}. Thus, the number of collisions at any two consecutive
pre-fractal generations are given as: 
\begin{equation}
\label{nc2}
\begin{tabular}{rccccc}
$\nu:$ \hspace{1cm} & $n_1$, & $n_2$, & $n_3$, & $\cdots$, & $n_n$\\
$\nu+1:$ \hspace{1cm} & $n_1,3n_1,n_1$, & $n_2,3n_2,n_2$, &
$n_3,3n_3,n_3$, & $\cdots$, & $n_n, 3n_n, n_n$
\end{tabular}
\end{equation}
A recursion relation for the collision numbers for $\nu+1$ generation
then can be written as
\begin{equation}
\label{seq}
(n_1,n_2,n_3)_{\nu+1}=n_\nu\times(1,3,1)
\end{equation}
where $n_\nu$ is the number of collisions of the trajectory from the
same part of the slit at the $\nu$th generation. 
Using the above recursion relation, the exact specular collision
numbers for $3^\nu$ trajectories at any generation $\nu$ can then be
obtained for $\phi=45^\circ$.

\section{Results and Discussions} 
The collision numbers and their spatial distribution along the slit
are computed for the $7$th and $8$th pre-fractal generation ($\nu=7$
and $8$). A number of $N_{tot}=3^\nu$ equally spaced particles are
launched from the slit. The spatial distribution of the collision
numbers is studied under two different conditions. Firstly, the
distribution of the specular collision numbers at $45^\circ$ angle of
incidence for which the exact collision numbers are available, is
investigated. Secondly, the effect of the surface micro-roughness
$(\Gamma)$ on the collision spectrum is studied for the same slit
angle. In this case, the collision spectrums are averaged over $160$
samples. The general quality of the numerical results is verified with
the exact collision spectrum for $\phi=45^\circ$ and $\Gamma=0^\circ$
obtained from the recursion relation (Eq. \ref{seq}).

\begin{figure}
\bigskip
\centerline{\psfig{file=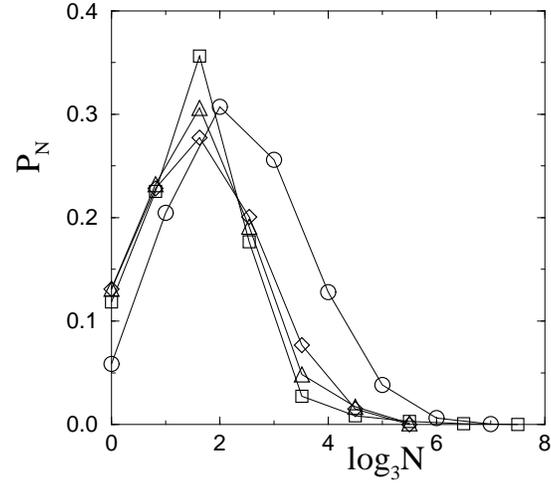,width=0.4\textwidth}}
\bigskip
\caption{ \label{pdist} Collision number distribution $P_N$ versus
 $N$. The distributions are given for $\Gamma=0$ (circles) specular
 reflection, $2^\circ$ (squares), $6^\circ$ (triangles), and
 $10^\circ$ (diamonds) at $45^\circ$ slit angle. The distribution
 shows that most of the particles make few collisions while a few
 trajectories exhibit a very large number of collisions. Note the
 existence of a tail in the distribution. }
\end{figure}

Before discussing the multifractality associated with the spatial
distribution of the collision numbers, it is important to consider their
statistical distributions. The probability to have a trajectory with
$N$ collisions is measured by
\begin{equation}
\label{prob}
P_N=\frac{n_N}{N_{tot}}
\end{equation}
where $n_N$ is the number of trajectories having $N$ collisions and
$N_{tot}$ is the total number of trajectories.  The distribution of
$P_N$ for different $\Gamma$ in Fig. \ref{pdist}. It can be observed
that most of the trajectories exhibit between $3$ and $9$
collisions. A large number of trajectories then leave the pre-fractal
surface making only a few collisions and only a few trajectories make
a large number of collisions. It is important to note that the
roughness $\Gamma$ has little effect on the collision number
distribution except the reduction in the maximum number of collisions
with increasing $\Gamma$. The total number of collisions is also very
weakly dependent on the randomness $\Gamma$. These features have
already been observed in the case of interaction of ballistic
trajectories with pre-fractal pores \cite{santra} in the Knudsen
diffusion regime where the collision number distribution was found to
be of the L\'evy type. However, up to the $8$th pre-fractal generation, 
no simple power law distribution of the
collision numbers was found.

\begin{figure}
\bigskip
\centerline{\hfill \psfig{file=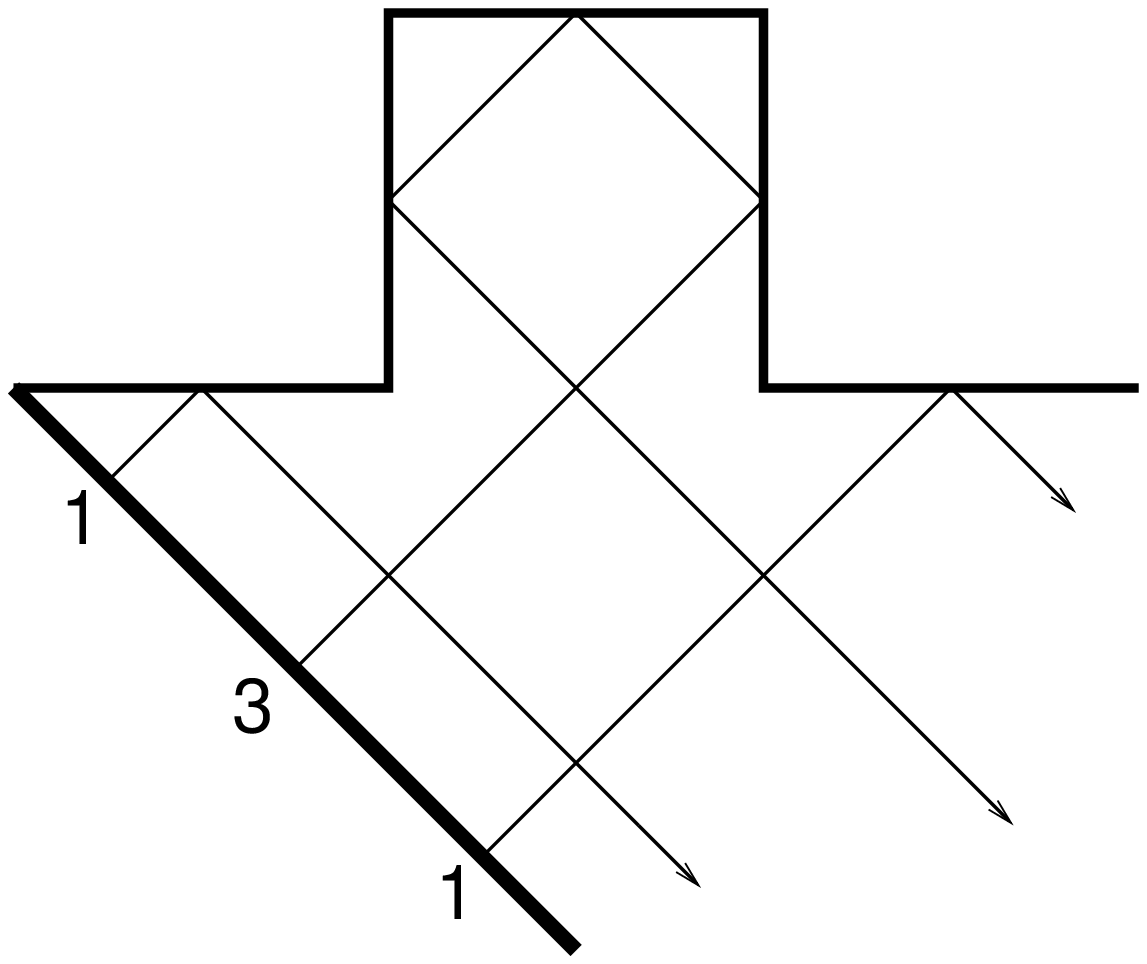,width=0.3\textwidth}
\hfill\hfill \psfig{file=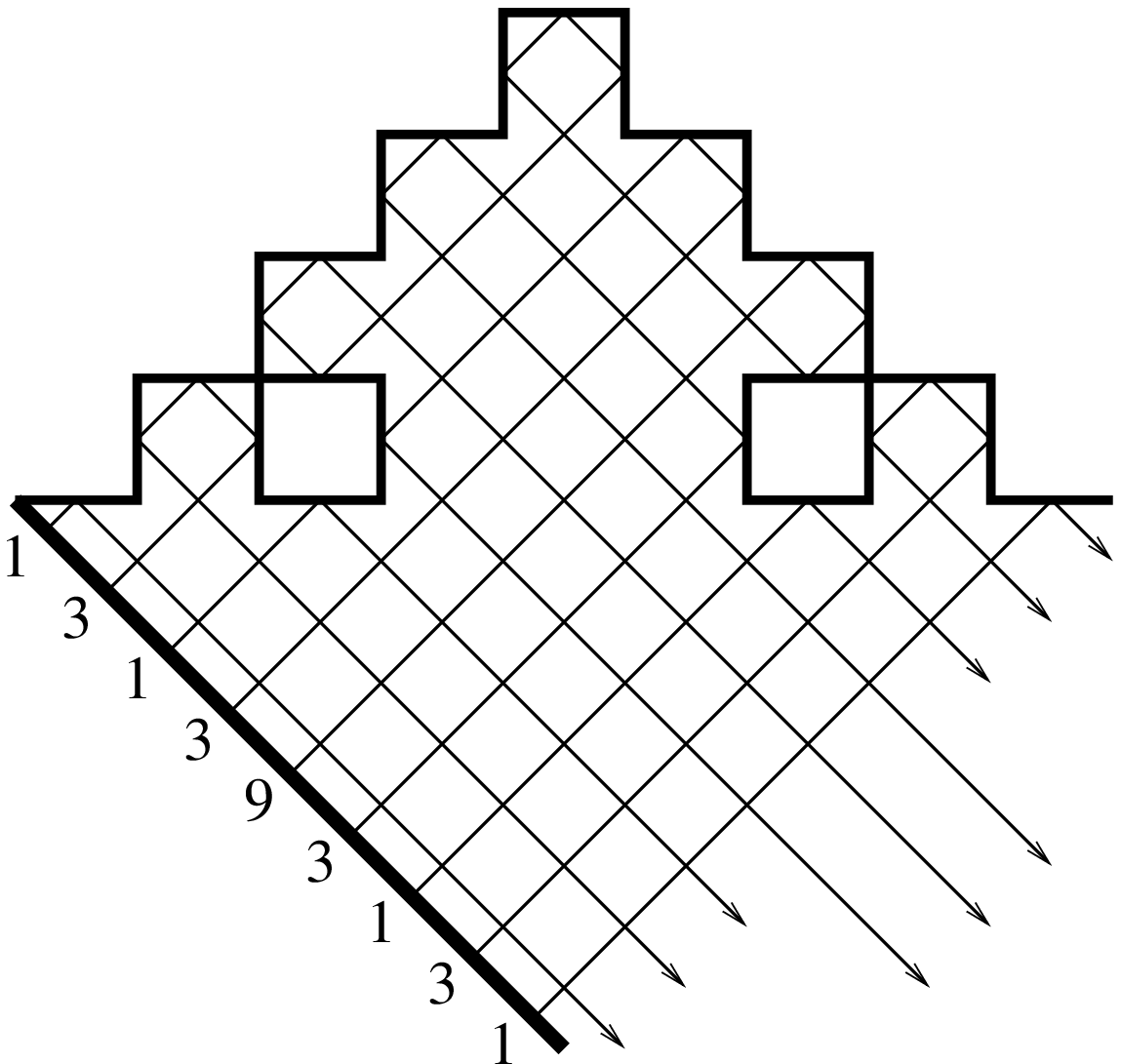,width=0.3\textwidth}\hfill}
\centerline{\hfill (a) \hfill\hfill (b)\hfill}
\bigskip
\caption{ \label{demoe} Multiplication of specular collision numbers
  from $\nu=1$ generation of the per-fractal $(a)$ to $\nu=2$
  generation of the per-fractal $(b)$ at $\phi=45^\circ$. Note that
  the collision numbers are uniformly distributed over the pre-fractal
  surface, each segment has one collision. }
\end{figure}

\begin{figure}
\bigskip
\centerline{\psfig{file=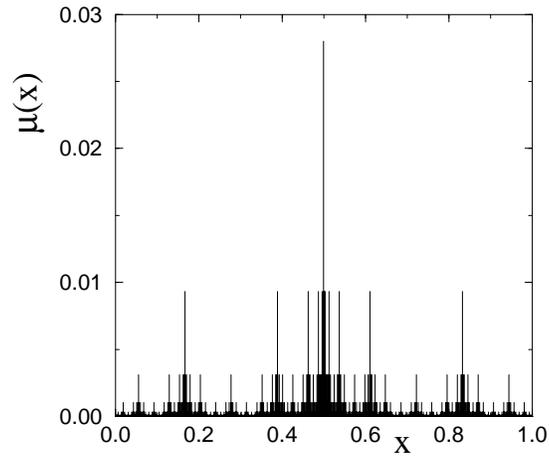,width=0.4\textwidth}}
\bigskip
\caption{\label{mux1} Plot of the relative collision numbers $\mu(x)$
versus $x$, initial position on the slit, for $\Gamma=0^\circ$ specular
reflections. This ``specular'' spectrum is symmetric around the central
position as it is expected. }
\end{figure}

To study the multifractal aspect of the trajectory distribution, a
collision spectrum is obtained. The collision spectrum is a
distribution of the relative collision numbers of the trajectories
over the slit. The relative collision number $\mu(x)$ of a particle
trajectory originating from a position $x$ on the slit and making
$n(x)$ collisions with the surface is given by
\begin{equation}
\label{mu}
\mu(x) =\frac{n(x)}{\sum_xn(x)}
\end{equation}
where the sum is over all the particles. This normalized collision
number is the measure under study here. In Fig. \ref{mux1}, $\mu(x)$
is plotted against the initial particle position keeping $\Gamma=0$
(specular reflection) for $\nu=7$. The value of $\mu(x)$ is
represented by impulses at the initial positions $x$. The spectrum is
symmetrical around the central position, as expected. Most interesting
to note, this spectrum corresponds to a uniform exploration of the
irregular surface as there is a single collision per segment (see
Fig. \ref{demoe}). The maximum number of collisions made by the
central trajectory is $3^7$ and the total number of collisions made by
all the particles is $5^7$ (the first generation has $5$ collisions)
at $\nu=7$. Thus, $\mu_{max}$ is given by $\mu_{max} = 3^7/5^7\approx
0.03$, as found numerically.

\begin{figure}
\bigskip
\centerline{\psfig{file=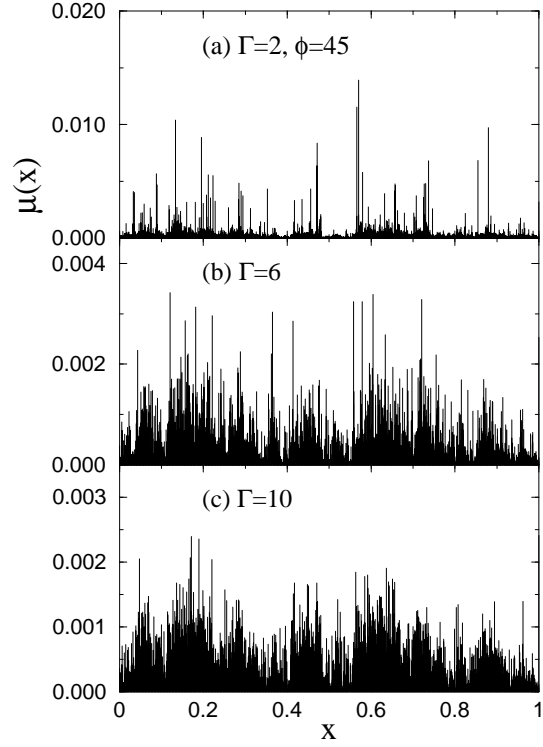,width=0.4\textwidth}}
\bigskip
\caption{\label{mux2} Plot of relative collision numbers $\mu(x)$ as a
function of the initial position ($x$) on the slit for ($a$)
$\Gamma=2^\circ$, ($b$) $\Gamma=6^\circ$, and ($c$)
$\Gamma=10^\circ$. Note the general decrease of the values and the
increased spreading of the distribution as $\Gamma$ increases. }
\end{figure}

The collision spectrums obtained in the presence of surface
micro-roughness, $\Gamma=2^\circ$ ($a$), $\Gamma=6^\circ$ ($b$), and
$\Gamma=10^\circ$ ($c$) are shown in Fig. \ref{mux2} for the $\nu=7$th
pre-fractal generation. Due to micro-roughness, the spectrums become
more uniform and the value of $\mu_{max}$ is decreased with respect to
the specular value. The value of $\mu_{max}$ is decreasing with
increasing $\Gamma$ as it is also seen in the collision number
distribution (Fig. \ref{pdist}) that the maximum number of collisions
is decreasing with $\Gamma$.

\begin{figure}
\bigskip
\centerline{\psfig{file=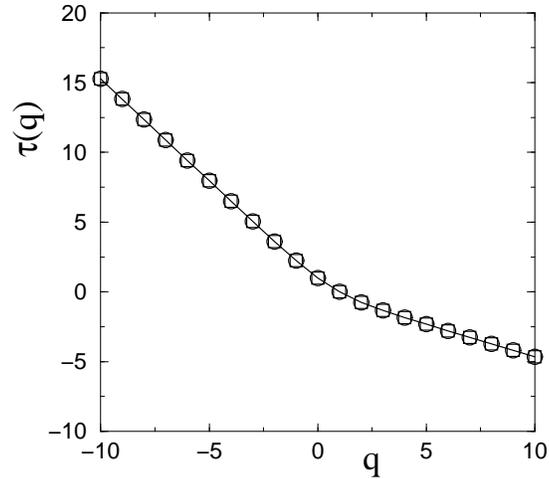,width=0.4\textwidth}}
\bigskip
\caption{\label{expo} Plot of the numerical $\tau(q)$ against $q$ for
different generations $\nu=7$ (circles) and $\nu=8$ (squares) with
$\Gamma=0$.  The collision exponents remains almost the same. The
solid line represents the theoretical $\tau_m(q)$ for the trinomial
mass distribution.}
\end{figure}

To show the multifractal nature of the distribution $\mu(x)$, it is
necessary to study the scaling of the $q$-moments of the measure over
different length scales on the slit. If the measure $\mu(x)$ is
multifractal and the slit is divided into $n_\epsilon$ boxes of size
$\epsilon$, then the weighted number of boxes $N(q$,$\epsilon)$ is
given by
\begin{equation}
\label{nep}
N(q,\epsilon)=\sum_{i=1}^{n_\epsilon} \mu_i^q \approx \epsilon^{-\tau(q)}
\end{equation}
where $\mu_i$ is the sum of the relative collision number of
trajectories initiated in the $i$th box. Here $\tau(q)$ is called the
``collision exponent''. In Fig. \ref{expo}, the values of $\tau(q)$,
obtained by the box counting method, are plotted against $q$ for
$\nu=7$ (circle) and $8$ (square) with $\Gamma=0$ (specular
reflections). The values of $\tau(q)$ do not depend on $\nu$, the
macro-irregularity of the surface. Note that, $\tau(0)$ is $\approx 1$
and $\tau(1)$ is $\approx 0$ here. $\tau(0)$ corresponds to the
dimension of the slit which is $1$ here and $\tau(1)$ is zero because
$\sum_x\mu(x) = 1$. The fact that a sequence of exponents are obtained
irrespective of the generation $\nu$ confirms the multifractal nature
of the spectrum. Note that the multifractality obtained for the
specular collision spectrum at $45^\circ$ angle of incidence
corresponds to a uniform exploration of the irregular surface
(Fig. \ref{demoe}). Consequently, a measure defined over the irregular
surface instead of the slit will not exhibit multifractality.

A check of the validity of our numerical computations can be made for
the case $\phi=45^\circ$ with specular reflections. In that case there
exists a trinomial multiplicative cascade \cite{evertsz-mandel}. A
comparison between the collision spectrum and the multiplicative
cascade is made here. At the zeroth generation of the cascade
$k=0$, a uniform distribution of unit mass over an interval $[0$,$1]$
is considered. In the next generation $k=1$, the unit mass is
distributed over three equal intervals $[0,1/3]$, $[1/3,2/3]$, and
$[2/3,1]$ with weighted probabilities $p_0=1/5$, $p_1=3/5$, and
$p_2=1/5$ similar to the collision number distribution over the slit
at $\phi=45^\circ$. At $k=7$ generation, the mass distribution will be
exactly the same as the collision number distribution over the slit
obtained for $\nu=7$ generation of the pre-fractal surface. The
measure over the unit interval remains always conserved in the
iteration process since $\sum p_i=1$. The mass exponent $\tau_m(q)$
can be calculated in terms of $p_i$ as
\begin{equation}
\label{taue}	
\tau_m(q)= \lim_{k\rightarrow \infty}
\frac{\ln[p_0^q+p_1^q+p_2^q]^k}{\ln3^{-k}} = -
\frac{\ln[p_0^q+p_1^q+p_2^q]}{\ln3}. 			
\end{equation}	
The solid line in Fig. \ref{expo} represents the mass exponent
$\tau_m(q)$ for this trinomial mass distribution. With no surprise,
there exists an excellent agreement between the exponents for the
specular collision number distribution at $\phi=45^\circ$ and the
trinomial mass distribution.
 
\begin{figure}
\bigskip
\centerline{\psfig{file=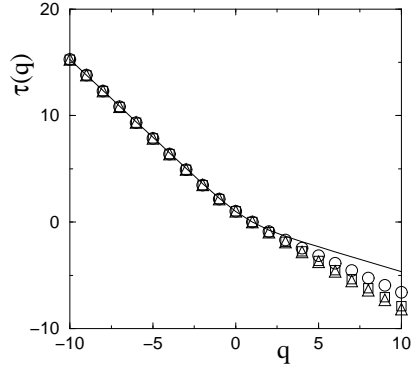,width=0.3\textwidth}}
\bigskip
\caption{\label{tqrg} Plot of $\tau(q)$ against $q$ for $\Gamma=2$
(circles), $\Gamma=6$ (squares), and $\Gamma=10$ (triangles) at
$45^\circ$ slit angle. The solid line represents $\tau(q)$ for the
specular collisions.  The randomness factor $\Gamma$ has a significant
effect on $\tau(q)$ for large positive $q$ values. }
\end{figure}

The effect of micro-roughness ($\Gamma$) on the collision exponents is
discussed now. In Fig. \ref{tqrg}, $\tau(q)$ is plotted against $q$
for $\Gamma=2^\circ$ (circle), $6^\circ$ (square), and $10^\circ$
(triangle). The solid line represents the collision exponents for
$\Gamma=0^\circ$, the specular reflections. At large positive $q$, the
micro-roughness $\Gamma$ has a significant effect on $\tau(q)$ with
respect to $\Gamma=0^\circ$. It is expected because the spectra change
with $\Gamma$. Again a sequence of collision exponents $\tau(q)$ is
obtained whatever the micro-roughness $\Gamma$. This means that the
spectrum are still of multifractal nature.

\begin{figure}
\bigskip
\centerline{\psfig{file=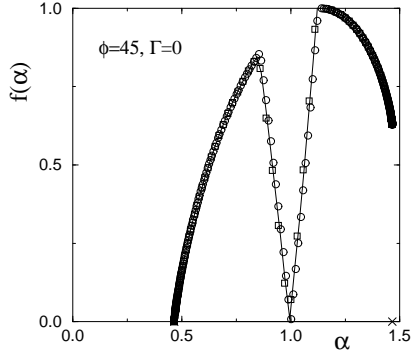,width=0.3\textwidth}}
\bigskip
\caption{\label{fa45} Plot of $f(\alpha)$ against the
Lipschitz-H$\ddot{o}$lder exponent $\alpha$ for specular collisions at
$\nu=7$ (circles) and $\nu=8$ (squares). The crosses represent the
values of $\alpha_{min}$ and $\alpha_{max}$ and the solid line
represents $f(\alpha)$ for trinomial mass distribution. The curves are
identical.}
\end{figure}

To determine the nature of the multifractality, the dimensions
$f(\alpha)$ are obtained as usual as a function of the
Lipschitz-H$\ddot{o}$lder exponent $\alpha$ through a Legendre
transformation \cite{evertsz-mandel}
\begin{equation}
\label{falfa}	
\alpha(q) = -\frac{d}{dq}\tau(q), \hspace{1cm} f(\alpha)=q\alpha(q)+
\tau(q)
\end{equation}	
of the sequence $\tau(q)$. The curve $f(\alpha)$ for specular
reflexions is discussed first. In Fig. \ref{fa45}, it is represented
by circles. The solid line represents $f(\alpha)$ for the trinomial
mass distribution. Naturally, there is a good agreement between
$f(\alpha)$ of specular collision distribution and the trinomial mass
distribution. Since the slit here is a one dimensional object, the
$f(\alpha)$ curves are always $\leq 1$ and $f_{max}(\alpha)$ is equal
to $1$. One observes two parabolas in the $f(\alpha)$ curve. This
means that the multifractal is of trinomial nature. The values of
$\alpha_{min}$ and $\alpha_{max}$ for specular reflections at
$\phi=45^\circ$ can be predicted. The increment in the measure over a
length $\xi$ to $\xi+\delta$ is $\mu_\xi = \mu(\xi+\delta) -
\mu(\xi)$. The Lipschitz-H$\ddot{o}$lder exponent $\alpha(\xi)$ is
defined as $\mu_\xi=\delta^{\alpha(\xi)}$. After $n$th iteration,
$\delta=1/3^{n}$ and $\alpha(\xi)$ is given by
\begin{equation}
\label{alxi}
\alpha(\xi)=\frac{\ln\mu_\xi}{\ln\delta}=-\frac{\frac{\xi}{2}\ln p_0
+(1-\xi)\ln(1-p_0-p_2)+\frac{\xi}{2}\ln p_2} {\ln3}
\end{equation}
where $p_0=p_2=1/5$ and $p_1=1-p_0-p_2=3/5$ for the case considered
here. $\alpha_{min}$ and $\alpha_{max}$ correspond to $\xi=0$ and
$\xi=1$ respectively and they are obtained as $\alpha_{min} =
-\ln(3/5)/\ln3 \approx 0.465$ and $\alpha_{max} = -\ln(1/5)/\ln3
\approx 1.465$. These two extreme values are marked by two crosses on
the $\alpha$-axis. It can be seen that $\alpha_{min}$ and
$\alpha_{max}$ obtained numerically for the specular collision
spectrum at $\phi=45^\circ$ are very close to the predicted
values. Note that, as expected, the whole $f(\alpha)$ versus $\alpha$
is independent of the generation $\nu$ of the pre-fractal surface. This is
shown in Fig. \ref{fa45}, where the data for $\nu=7$ (circles) and
$\nu=8$ (squares) collapse onto the same curve. It is also interesting
to notice that $\alpha_{max}$ is exactly equal to the dimension
$\ln5/\ln3$ of the pre-fractal and $\alpha_{min} = \alpha_{max} -
1$. These features are specific of the prefractal geometry and
the angle of incidence considered here.

Always in the case of specular reflections, but for other angles of
incidence like $\phi=30^\circ$ and $60^\circ$, it is found that the
$f(\alpha)$ versus $\alpha$ curves also consist of two
parabolas. However, in both cases, the values of $\alpha_{min}$ and
$\alpha_{max}$ keep on changing with the generation $\nu$ upto $8$th
generation. This is attributed to the fact that the multiplicative
process in these cases is more involved than in the case of $45^\circ$
incidence. There should exist for these cases a cascade of different
sequential values of $p_0$, $p_1$ and $p_2$ so that the asymptotic
multifractal measure is still not reached for $\nu = 8$.

\begin{figure}
\bigskip
\centerline{\psfig{file=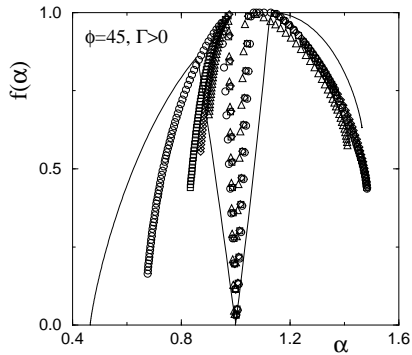,width=0.3\textwidth}}
\bigskip
\caption{\label{fadf} Plot of $f(\alpha)$ against $\alpha$ for
 different $\Gamma$ at $\phi=45^\circ$: circles for $\Gamma=2^\circ$,
 squares for $\Gamma=6^\circ$ and diamonds for $\Gamma=10^\circ$.
 Triangles correspond to the totally diffusing surface and the solid
 line corresponds to $f(\alpha)$ of the trinomial mass distribution. }
\end{figure}

The effect of micro-roughness of the surface on the $f(\alpha)$ curve
is discussed now. In Fig. \ref{fadf}, $f(\alpha)$ versus $\alpha$
curves are plotted for $\Gamma=2^\circ$ (circles), $6^\circ$
(squares), and $10^\circ$ (diamonds) at $\phi=45^\circ$ and
$\nu=7$. The solid line represents the $f(\alpha)$ curve for specular
collisions. The triangles represent the results for an hypothetic
``totally'' diffusing surface for which re-emission would be isotropic,
$i.e.$; occuring randomly in any direction between $-90^\circ$ to
$90^\circ$ with respect to the normal, independent of the angle of
incidence. This is equivalent to the $\Gamma\rightarrow 90^\circ$
limit. There are two important things to note. First, the
micro-roughness (or random character of the particle re-emission) of
the surface is unable to destroy either the multifractality or the
trinomial nature of the multifractality even at the maximum possible
value of $\Gamma$, corresponding to the ``totally diffusing
surface''. Second, the values of $\alpha_{min}$ depends strongly on the
degree of surface roughness $\Gamma$ whereas $\alpha_{max}$ remains
almost constant. The value of $\alpha_{min}$ increases as a function
of $\Gamma$. This is expected. As $\Gamma$ increases, the value of
$\mu_{max}$ decreases, the spectrum becoming more uniform as seen in
Fig. \ref{mux2} and the smaller value of $\mu_{max}$ corresponds to
the larger value of $\alpha_{min}$. The value of $\alpha_{max}$ could
also be estimated from the knowledge of the minimum measure
$\mu_{min}$. For a single realization, the total number of collisions
is $\approx 10^5$. Thus, $\mu_{min} \approx 1/10^5$ (for a trajectory
having a single collision) and the value of $\alpha_{max}$ is then
given by $\ln\mu_{min}/\ln\delta = \ln(1/10^5)/\ln(1/3^7) \approx
1.497$ very close to what is obtained here. Since the value of
$\mu_{min}$ is almost independent of the surface roughness (total
number of collisions almost independent of $\Gamma$) in this collision
process, the value of $\alpha_{max}$ remains unchanged.

From these results, it is found that the multifractality as well as
the trinomial nature of the multifractality associated with this
collision process remains unaltered irrespective of the surface
micro-roughness. This is because, the multifractality of the collision
spectrum is the outcome of a multiplicative cascade of the spatial
sub-distributions which are independent of the various situations
considered here. The trinomial nature of the multifractal is due to
the type of pre-fractal chosen here to model the surface
macro-irregularity: a quadratic Koch curve, each element being divided
into three elements in the next generation. A different pre-fractal
generator, $i.e.$; a different model of surface macro-irregularity
will then lead to a different multifractal behavior.

The information dimension \cite{grass} is the value of $f(\alpha)$
when $f(\alpha) = \alpha$. The information dimension $D_I$ is found to
be $D_I\approx 0.854$ for $45^\circ$ angle of incidence with
$\Gamma=0$ and at $\nu=7$. This indicates that the measure is
supported by a minor fraction of the particle trajectories. This also
illustrates the fact that most of the trajectories interacts weakly
with the fractal surface. Interestingly, the information dimension
increases with the random character of the particle-surface
interaction. Indeed for $\Gamma=2^\circ$, $6^\circ$ and $10^\circ$ the
values of information dimensions are $D_I\approx 0.877$, $0.934$ and
$0.952$ respectively. For the diffused surface $D_I\approx
0.964$. This corresponds to more and more uniform collision number
distribution on the slit.
 
It is interesting to comment on the uniform exploration of the
surface, each element receiving the same number of particles at least
for an angle of incidence of 45 degrees.  This is analogous to the
results obtained by Andrade et al \cite{jose} through a molecular
dynamics study of Knudsen diffusion in the same geometry. In this last
case, it is the first collision that is uniformly distributed over the
perfectly absorbing surface whereas here, the surface is perfectly
reflecting.  The uniform exploration found in these both cases
constitutes a qualitative evidence of the absence of screening in
Knudsen diffusion. It can be considered as a remarkable fact that,
seen from the source of particles, the collision numbers are strongly
non-uniform, while seen from the surface the exploration is
uniform. This may have interesting consequences for catalyst
deactivation which should also be uniformly distributed.

\section{Conclusion}
The collision number distribution and their spatial behaviour are
obtained for $2D$ ballistic particles interacting with a pre-fractal
curve of dimension $\ln5/\ln3$. The trajectories considered here
correspond to the Knudsen diffusion regime. The micro-roughness of the
surface, or the random character of the particle-surface interaction,
is included through a randomness parameter $\Gamma$ in the reflection
angle. It is found that most of the particles make only a few
collisions whereas a few trajectories exhibit a large number of
collisions. A multifractal measure of the spectrum is defined in terms
of the relative collision number over a linear slit from which the
trajectories are launched. A sequence of ``collision exponents''
$\tau(q)$, is determined by the box counting method and it is found
that the sequence depends on the moment $q$. Thus, the spatial
distribution of the collision numbers is a multifractal spectrum. The
fractal dimensions $f(\alpha)$ depend on the random character of the
particle re-emission $(\Gamma)$. Irrespective of the values of
$\Gamma$, the plot of $f(\alpha)$ versus Lipschitz-H$\ddot{o}$lder
exponent $\alpha$ consist of two parabolas. The spectrum is then that
of a multiplicative trinomial multifractal independent of surface
micro-roughness. The information dimension $D_I$ of the measure is
found to be smaller than $1$ and increases with the random character
of the particle surface interaction. Different type of surface
macro-irregularity would lead to different multifractal
behavior. Notably, the surface however works uniformly, indicating an
absence of screening in the Knusdsen diffusion regime.

\newpage

\begin{table}
\begin{tabular}{c}
\hline Number of Collisions\\ \hline
\begin{tabular}{p{1.5cm}ccp{0.5cm}ccp{0.5cm}ccp{0.5cm}ccp{0.5cm}cc
p{0.5cm}ccp{0.5cm}ccp{0.5cm}ccp{0.5cm}ccc} 
$\nu=0$ &&&&&&&&&&&&&& $1$ &&&&&&&&&&&&&\\ 
$\nu=1$ &&&&& $1$ &&&&&&&&& $3$ &&&&&&&&& $1$ &&& \\ 
$\nu=2$ && $1$ &&& $3$ &&& $1$ &&& $3$ &&& $9$ &&& $3$ &&& $1$ &&& $3$
  &&& $1$ &\\ 
$\nu=3$ & $1$ & $3$ & $1$ & $3$ & $9$ & $3$ & $1$ & $3$ & $1$ & $3$ & $9$ &
$3$ & $9$ & $27$ & $9$ & $3$ & $9$ & $3$ & $1$ & $3$ & $1$ & $3$ & $9$
  & $3$ & $1$ & $3$ & $1$\\
\hline
\end{tabular}
\end{tabular}
\caption{\label{table} Number of specular collisions made by ballistic
  trajectories incident at $45^\circ$ and interacting with a quadratic
  Koch curve of dimension $\ln5/\ln3$ at different generations
  $\nu$. The number of collisions at a generation $\nu +1 $ is given
  by $n_\nu\times(1,3,1)$ where $n_\nu$ is the number of collisions in
  the previous generation for the same trajectory.}
\end{table}

\newpage

\end{document}